\documentstyle[12pt,twoside,fleqn,espcrc1,epsfig]{article}
\newcommand{\be}{\begin{eqnarray}}
\newcommand{\ee}{\end{eqnarray}}

\newcommand{\AmS}{{\protect\the\textfont2
  A\kern-.1667em\lower.5ex\hbox{M}\kern-.125emS}}

\title{Coulomb distortion of $\pi^+/\pi^-$ as a tool
       to determine the fireball radius in central 
       high energy heavy ion collisions\thanks{Support 
                received in part by CONACyT 
                M\'exico under grant I27212-E, by the U.S.
                National Science Foundation under grant NSF PHY94-21309 and
                by the U.S. Department of Energy under grants 
                DE-FG02-87ER40328, DE-AC0376SF00098 and DE-FG03-93ER40792.}}
\author{Alejandro Ayala\address{Instituto de Ciencias Nucleares\\
        Universidad Nacional Aut\'onoma de M\'exico\\
        Apartado Postal 70-543 M\'exico D.F. 04510, M\'exico.}%
        ~Sangyong Jeon\address{Nuclear Science Division, Lawrence Berkeley
        Laboratory\\
        Berkeley, California 94720, USA.}%
        ~and
        Joseph Kapusta\address{School of Physics and Astronomy\\
        University of Minnesota\\
        Minneapolis, Minnesota 55455, USA.}}

\begin{document}

\maketitle

\begin{abstract}

We compute the Coulomb distortion produced by an expanding and highly charged
fireball on the spectra of low transverse momenta and mid rapidity pions 
produced in central high energy heavy ion collisions. We compare to data on 
Au+Au at $11.6A$ GeV from E866 at the BNL AGS and of Pb+Pb at $158A$ GeV from 
NA44 at the CERN SPS. We match the fireball expansion velocity with the average
transverse momentum of protons and find a best fit to the charged pion ratio
when the fireball radius is about 10 fm at freeze-out. This value is common
to both AGS and SPS data. 

\end{abstract}

\vspace{0.2in}
An important feature to account for in the analysis of the spectra of 
secondaries produced in the collision of heavy systems is the presence
of a large amount of electric charge. Due to the long-range nature of the
electromagnetic interaction, the spectrum of charged particles will be 
distorted even after freeze-out. For central collisions, this Coulomb effect 
can be more significant when there is strong stopping and the participant 
charge in the central rapidity region is an important fraction of the initial
charge. 

Another feature to consider is that the field-producing charge distribution 
is in general not static but rather participates in the dynamics
responsible for matter expansion after the collision. The combined role played 
by Coulomb distortions and expansion in the description of charged particle 
spectra has been the subject of some recent work. Koch~\cite{Koch} and Barz, 
Bondorf, Gaardho$\!\!\! /$je and Heiselberg~\cite{Barz} have developed
approximate models to describe the situation in which expansion takes place 
predominantly along the collision axis. In this work, we focus on the 
description of Coulomb effects on pion spectra from a spherically symmetric 
expanding source. We compare our calculation to mid-rapidity pions produced in
central Au+Au reactions at $11.6A$ GeV from E866 at the BNL AGS and in central
Pb+Pb reactions at $158A$ GeV from NA44 at the CERN SPS. A detailed analysis 
can be found in Refs.~\cite{Ayala}.

Before proceeding on to the model calculation, let us say a few words about 
the assumption of spherical symmetry. It has been known for some time
that the momentum distributions of secondaries are somewhat forward-backward 
peaked, specially at the SPS, even for central collisions~\cite{KLM-NA49}, and
this observation can cast some doubt about the validity of a model that
assumes a spherically symmetric fireball. However, spherical geometry is not 
essential to the basic physics and can be relaxed at the expense of additional
computing time. Nevertheless, let us provide the following arguments in favor 
of its use. First, we will be comparing with the transverse momentum 
distributions at mid-rapidity where the impact of spherical asymmetry should 
be less important than near the fragmentation regions. Second, pion 
interferometry of central Au+Au collisions at the AGS~\cite{Barrette} and of 
central Pb+Pb collisions at the SPS~\cite{Wiedemann} both yield comparable 
values for the transverse and longitudinal radii at the time of pion
freeze-out. These radii are about twice the radius of a cold gold or lead 
nucleus. Third, as we will later see, the transverse surface of the 
fireball needs to expand with a speed more than 90\% that of light in 
order to reproduce the average proton transverse momentum.  Since the 
longitudinal surface of the fireball cannot travel faster than the speed of 
light, this means that in velocity space the fireball is 
nearly symmetric. These phenomena, although not yet measured at that time, 
were already known to Landau~\cite{Landau}. The essential insight from his 
model is not the degree of stopping but rather the point that significant 
transverse expansion sets in after the longitudinal and transverse radii 
become comparable in magnitude. Thereafter, from the point of view of a 
distant observer, the expansion is not as asymmetric as one might originally 
think. This model was later developed by others, including Cooper, Frye and 
Schonberg~\cite{Cooper}. 

A uniformly charged sphere which has a total charge $Ze$ and whose radius $R$
increases linearly with time $t$ from a value $R_0$ at time $t_0$ at a
constant surface speed $v_s$ produces an electric potential
\be
  V(r,t)=\left\{ \begin{array}{ll}
  Ze/4\pi r \,\,,   &   r \geq R = v_st \\
  Ze(3R^2-r^2)/8\pi R^3 \,\,,   &   r \leq R = v_st
  \end{array}
  \right. \, .
  \label{eq:potential}
\ee
In the center-of-mass frame of the fireball the charge moves radially 
outwards, hence there is no preferred direction and consequently the magnetic 
field produced by this moving charge configuration vanishes. The fireball 
parameters are related by $R_0=v_st_0$. If $f^{\pm}({\bf r},{\bf p},t)$ 
represents the $\pm e$ test particle phase space distribution then, when 
ignoring particle collisions after decoupling, its dynamics is governed by
Vlasov's equation.
\be
   \left[ \frac{\partial}{\partial t} +
   \frac{{\bf p}}{E_p}\cdot\nabla_r \pm
   e{\bf E}({\bf r},t)\cdot\nabla_p \right]
   f^{\pm}({\bf r},{\bf p},t)=0\, ,
   \label{eq:Vlasov}
\ee
where $E_p=\sqrt{p^2+m^2}$, $m$ is the meson's mass and 
${\bf E}({\bf r},t) = -\nabla_r V(r,t)$ is the time-dependent electric field 
corresponding to the potential $V(r,t)$.

The solution is found by the method of characteristics. This involves solving 
the classical equations of motion and using the solutions to evolve the 
initial distribution, taken to be thermal
\be
   f^{\pm}({\bf r},{\bf p},t_0) =
   \exp\left\{ -\left( E_p \pm V(r,t_0) \right)/T \right\}\, ,
\ee
forward in time. The pion's asymptotic momentum is computed numerically by a 
sixth order Runge-Kutta method with adaptive step sizes from a set of initial
phase-space positions. The final momentum distribution is the result of 
computing the trajectories for many initial phase space points. The initial 
radial position was incremented in N$_r$=50 steps with spacing 
$\Delta r/R_0$ = 0.02. The initial momentum was incremented in N$_p$=600 
steps with spacing $\Delta p$ = 1 MeV. The cosine of the angle between the 
initial position and momentum vectors was incremented in N$_z$=100 steps of 
size $\Delta \cos\theta=0.02$.  Hence the total number of trajectories 
computed was $3\times 10^6$ for each set of initial conditions.

To determine the value of the surface expansion velocity, we match the
average transverse momentum, as inferred from our assumption of a uniformly
expanding sphere, with the measured one. The result is
\be
   \langle p_T \rangle = \pi \, \frac{2-(2+v_s^2)
   \sqrt{1-v_s^2}}{4v_s^3} \, m_P \, .
   \label{suex}
\ee
Note that $v_s$ should {\it not} be interpreted as a hydrodynamic flow 
velocity, rather, it embodies the combined effects of hydrodynamic flow and
thermal motion of the net charge carriers, mainly protons.

For central Pb+Pb collisions at 158$A$ GeV~\cite{NA44-NA49}, the number of 
protons per unit rapidity in the central region is on the order of 30. If we 
consider that the rapidity region spanned by the fireball is between 1 and 5 
units, then we take as the effective fireball's charge $Z=120$. To model
the primordial distribution, we use an exponential parametrization of
$dN/p^2dp$ from which we extract an effective temperature $T_{eff}= 110$ 
MeV up to $m_T-m=500$ MeV. Figure~1 shows this representation in comparison
to the transverse mass distributions of positive and negative pions. Both NA44
and NA49~\cite{NA44,NA49} have reported the transverse mass distribution in 
central Pb+Pb collisions at mid-rapidity to be 
$dN/p_Tdp_T \propto \exp(-m_T/T_P)$ with $T_P$ = 290 MeV. This corresponds 
to an average transverse momentum of 825 MeV. The value $v_s$ = 0.916 gives, 
according to Eq.~(\ref{suex}), the match of the model to the proton spectra. 
The best fit to the $\pi^+/\pi^-$ ratio is obtained for a value of the 
freeze-out radius for pions of 10 fm with a $\chi^2_{p.d.o.f.}=1.67$. 

\begin{figure}[h]
\vspace{-0.8in}
\begin{minipage}{75mm}
\epsfig{file=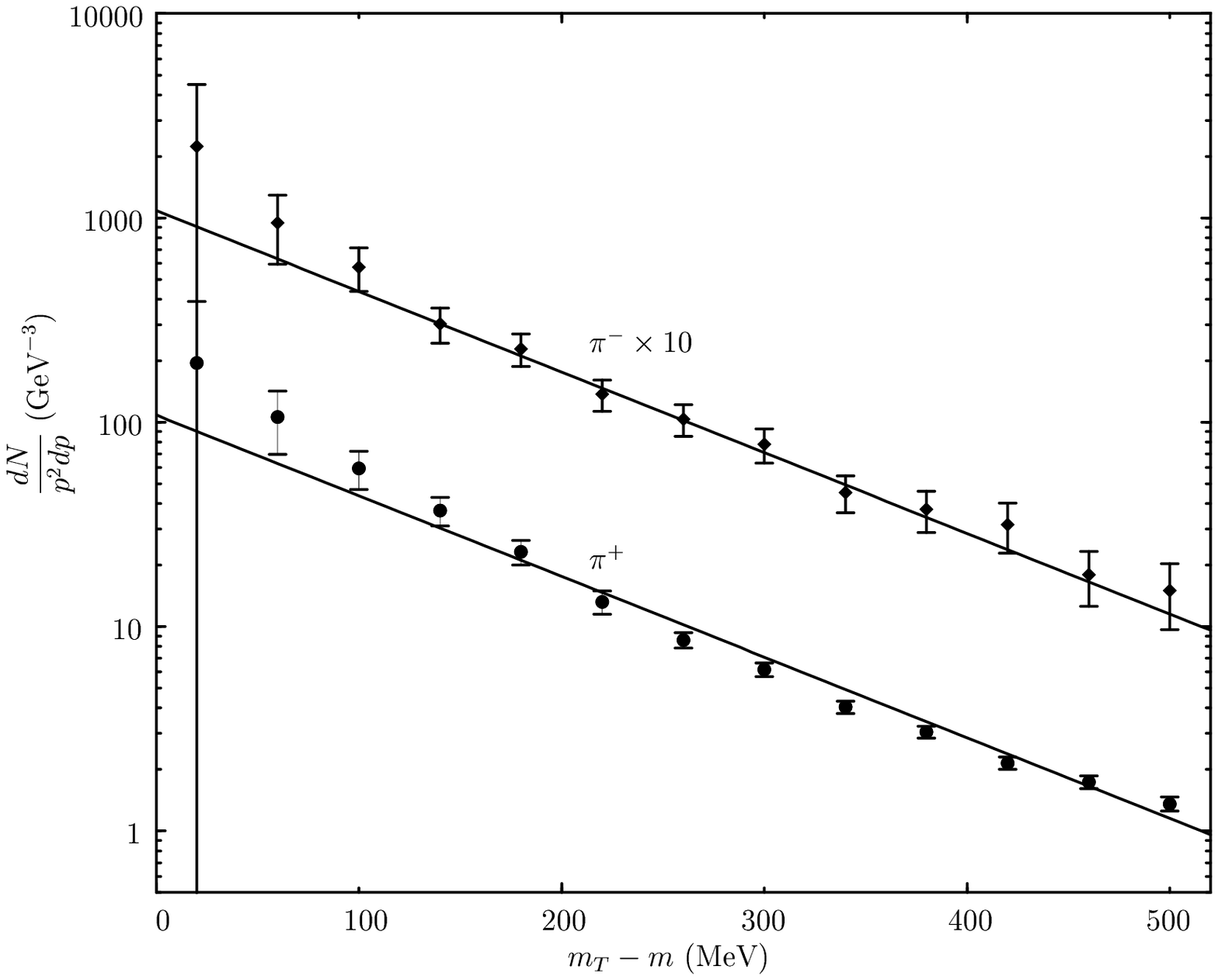,height=5.9in,width=3.4in}
\vspace{-2.8in}
\caption{ The transverse mass distributions at mid-rapidity
          for $\pi^+$ and $\pi^-$.  The solid lines are exponentials
          with an inverse slope of 110 MeV.  The data are from 
          NA44~\cite{NA44}. }
\end{minipage}
\hspace{\fill}
\begin{minipage}{75mm}
\epsfig{file=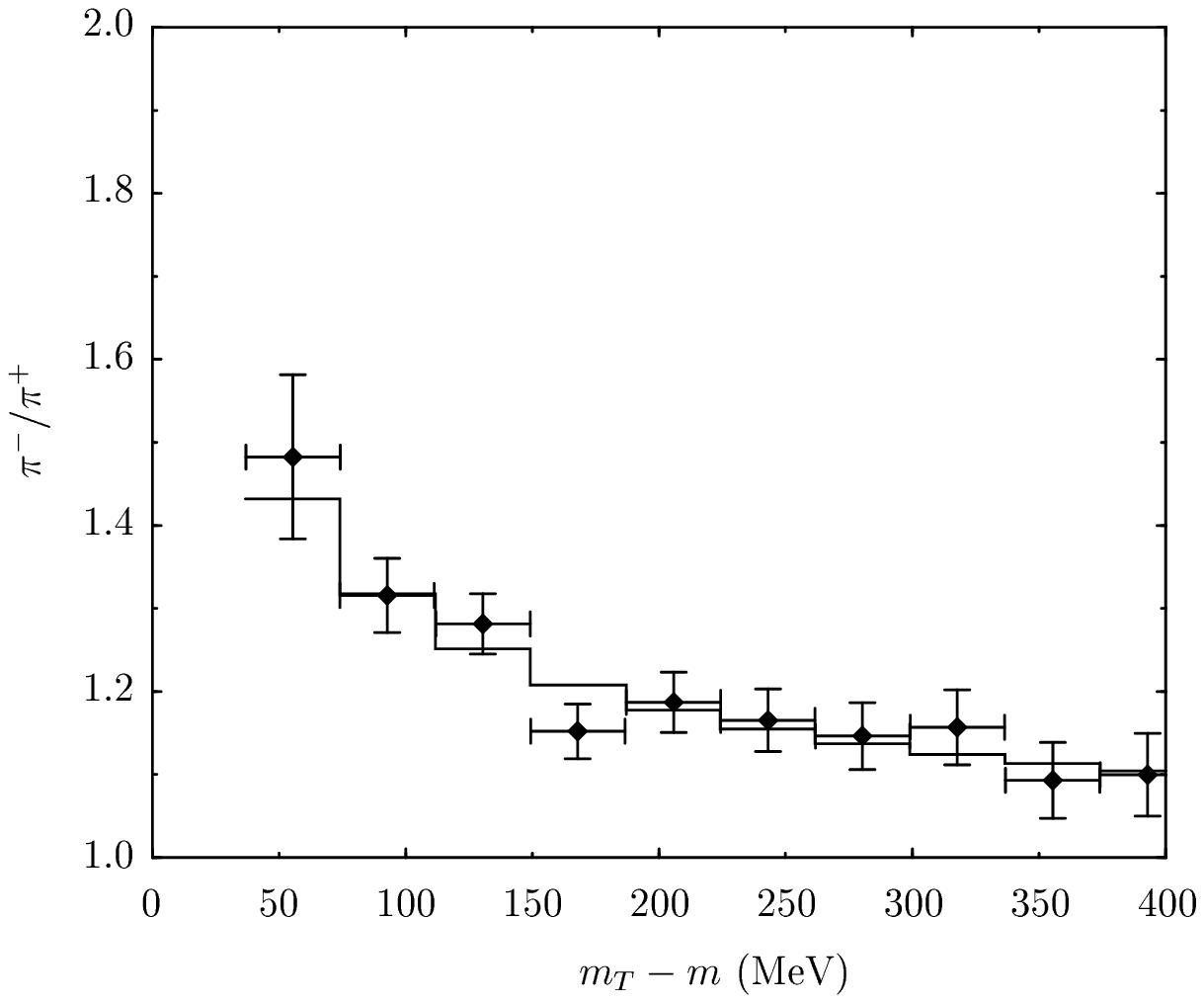,height=8.0in,width=4.27in}
\vspace{-4.5in}
\caption{ The ratio $\pi^-/\pi^+$ vs. $m_T - m$ at the AGS. The data is
          from Ref.~\cite{E866}. The fireball radius at pion freeze-out is 
          10 fm. }
\end{minipage}
\vspace{-0.2in}
\end{figure}

For central Au+Au collisions at 11.6$A$ GeV~\cite{E866}, the number of
participating protons may be estimated from the data to be $Z=116$. The average
proton transverse momentum is 820 MeV yielding the surface speed $v_s = 0.914$.
The slopes of the $\pi^+$ and $\pi^-$ distributions up to $m_T-m= 400$ MeV
are the same to within several MeV; their average is $111$ MeV. All of these 
quantities are remarkably similar to those in central Pb+Pb collisions at the 
much higher energy of the SPS. The best fit to the $\pi^-/\pi^+$ ratio is 
obtained for a value of the freeze-out radius for pions of 10 fm with a 
$\chi^2_{p.d.o.f.}=0.53$. The computed ratio is compared to the data in Fig.~2 
and its appearance is quite satisfactory.

In conclusion, we have shown that the suppression of the ratio $\pi^+/\pi^-$
in central Pb+Pb collisions at the SPS, as observed by NA44, and in central 
Au+Au collisions at the AGS, as observed by E866, can be quantitatively 
understood as a Coulomb effect generated by the electric field of an expanding
and highly charged fireball. This ratio provides a good measure of the size of
the fireball at decoupling. In principle, a different parametrization of the
primordial distribution, such as a two temperature fit, might lead to an
even better representation of data.

\end{document}